\documentstyle[11pt]{article}
\def\be {\begin{equation}}
\def\ee {\end{equation}}
\def\bea{\begin{eqnarray}}
\def\eea{\end{eqnarray}}
\def\x{\times}
\def\tfrac#1#2{{\textstyle{#1\over #2}}}
\begin{document}
\renewcommand{\arraystretch}{1.5}
\rightline{UG-3/97}
\rightline{hep-th/9703124}
\rightline{March 1997}
\vspace{1.5truecm}
\centerline{{\bf INTERSECTING BRANES AND SUPERSYMMETRY}\footnote{Presented at 
{\sl Supersymmetry and Quantum Field Theory},  
International Seminar dedicated to the memory of D.~V.~Volkov,
Kharkov State University (Kharkov, Ukraine), January 5-7, 1997.}}
\vspace{1.0truecm}
\centerline{\bf M.~de Roo}
\vspace{.4truecm}
\centerline{{\it Institute for Theoretical Physics}}
\centerline{{\it Nijenborgh 4, 9747 AG Groningen}}
\centerline{{\it The Netherlands}}
\vspace{2truecm}
\centerline{ABSTRACT}
\vspace{.5truecm}

We consider intersecting $M$-brane solutions of supergravity 
 in eleven dimensions. Supersymmetry
 turns out to be a powerful tool in obtaining such solutions 
 and their generalizations.

\vspace{.5truecm}

\noindent{\bf 1. Introduction}
\vspace{3mm}

The revival of the concept of strong-weak coupling duality has
 drastically changed our view of string theories. The five
 apparently different ten-dimensional superstring theories 
 are now interpreted as different limits of a single theory,
 the conjectured $M$-theory. The study of extended objects,
 which by duality must manifest themselves in each of the
 descendents of $M$-theory, has been  a decisive factor in
 establishing this picture of a united string 
 theory\footnote{For a recent review
 of these developments, see, e.g., \cite{Towns1}}.

Of particular interest are those extended objects ($p$-branes, where 
 $p$ is the dimension of the spatial extension) which satisfy
 a BPS-bound and preserve partial supersymmetry. 
 Such objects can satisfy a ``no-force'' condition,
 implying that static configurations
 of several such objects
 can exist due to a cancellation of the gravitational and
 gauge forces between them. Several authors have contributed
 to the rather complete picture that now exists of these
 intersecting $p$-brane configurations
 \cite{Papa1,Tseyt1,BeBeJa,Gaunt,Tseyt2,Costa,mult}. Here
 I would like to report on the work done in \cite{mult}, where
 a classification of multiple intersections in $D=10$ and $D=11$
 was obtained. I will limit myself to our results in 
 eleven dimensions, and, in
 the spirit of this meeting, I would like to discuss in
 particular how supersymmetry can be helpful in obtaining
 intersections of $M$-branes. In particular, we will find that
 supersymmetry is a useful guide in constructing the intersections
 of the $M2$- and $M5$-brane, and it shows that these should
 be extended to include objects with 1, 6, and 9 spatial extensions.

\vspace{3mm} 
\noindent{\bf 2.\ Pair Intersections}
\vspace{3mm} 

The basic solutions in $D=11$ are the $M2$-brane \cite{Duff-Stelle}: 
\begin{eqnarray}
\label{M2}
 ds^2 &=& {H^{-2/3}}\  dx^2_{(0-2)} - 
 {H_2^{1/3}} \ dx^2_{(3-10)}, \qquad
 F_{012i} = \partial_i H^{-1}, 
\end{eqnarray}
\noindent
where $H$ is harmonic on the eight-dimensional space transverse to the
 membrane,
 and the $M5$-brane solution\footnote{Supergravity in $D=11$ is formulated 
 in terms of a three-form gauge field. For the solutions considered
 here the contribution of the Chern-Simons term to the equations
 of motion, which depends on the three-form gauge
 field, does not contribute. In that case it is possible to
 represent the fivebrane in terms of 
 a six-form gauge field, the field strength $F_{012345i}$ being the dual of
 $F_{jklm}$.}:
 \cite{Guven}:
\begin{eqnarray}
\label{M5}
ds^2 &=& {H^{-1/3}}\  dx^2_{(0-5)} - 
 {H^{2/3}} \ dx^2_{(6-10)}, \qquad
 F_{012345i}= \partial_i H^{-1}.
\end{eqnarray}
\noindent
In this case  $H$ is harmonic on the five-dimensional transverse space.

For our purposes it is useful to represent the metric for these solutions
 pictorially as
\begin{equation}
\label{notation}
ds^2=\underbrace{\x \ \x \ ... \ \x}_{p+1}\  
        \overbrace{- \ - \ ... \ -}^{10-p}\,,
\end{equation}
where $\x$
 indicates a worldvolume coordinate of, $-$ a
 direction transverse to the $p$-brane. In this notation, the
 basic intersections \cite{Papa1,Tseyt1,Gaunt} 
 of the $M2$- and $M5$-brane can be represented 
 by\footnote{We denote the intersection of a $p_1$- and a $p_2$-brane
 over a common $q+1$ dimensional spacetime by $(q|p_1,p_2)$.}
\bea 
(0|M2,M2) &=& 
  {\scriptsize
   \left\{
     \begin{array}{c|cccccccccc}
       \x& \x& \x&  -&  -&  -&  -&  -&  -&  -& - \\
       \x&  -&  -& \x& \x&  -&  -&  -&  -&  -& -
     \end{array}\right.}\,, \label{M2M2} \\
 (1|M2,M5) &=&
 {\scriptsize
   \left\{ 
      \begin{array}{c|cccccccccc}
         \x & \x & \x &  - &  - &  - &  - & - & -  & -  & - \\
         \x & \x &  - & \x & \x & \x & \x & - & -  & -  & -             
      \end{array} \right.}\,,   \label{M2M5} \\
(3|M5,M5) &=&
 {\scriptsize
  \left\{
   \begin{array}{c|cccccccccc}
     \x& \x& \x& \x& \x& \x&  -&  -&  -&  -& - \\
     \x& \x& \x& \x&  -&  -& \x& \x&  -&  -& -
   \end{array}\right.}\,, \label{M5M54} \\
(1|M5,M5) &=& {\scriptsize\left\{
\begin{array}{c|cccccccccc}
\x& \x& \x& \x& \x& \x&  -&  -&  -&  -& - \\
\x& \x&  -&  -&  -&  -& \x& \x&  \x& \x& -
\end{array}\right.}\,. \label{M5M58}
\eea
Each intersection is determined by two harmonic functions, $H_1$ and $H_2$.
 We distinguish between overall worldvolume directions (both rows have an
 $\x$, the harmonic functions are in all cases 
 independent of these directions),
 relative transverse directions (only one row has an $\x$), and
 overall transverse directions (both rows have a $-$). 
 In (\ref{M2M2}-\ref{M5M54})
 either both $H_i$ must depend on the overall transverse
 directions, or one $H$ must depend on overall transverse, 
 the other on relative
 transverse directions. In (\ref{M5M58}) the dependence of the $H_i$ must
 be on the relative transverse directions only.

The metric for these basic pairs is easily constructed. In general,
 in the intersection of type $(q|q+r,q+s)$ the form of the metric is
\begin{eqnarray}
\label{general}
 ds^2 &=& H_1^{\alpha_1}H_2^{\alpha_2}\big\{
     dx^2_{(0-q)} - H_1 dx^2_{(q+1,q+s)}   \nonumber \\
 &&\qquad     H_2    
 dx^2_{(q+s+1,q+s+r)} -  H_{1}H_{2} dx^2_{(q+r+s+1,10)} \big\}\,.
\end{eqnarray}
Here $\alpha$ is $-2/3$ for $M2$, $-1/3$ for $M5$.
 The curvature tensors $F$ for the basic pairs correspond to
 the sum of the curvatures of the separate branes, except for
 (\ref{M5M58}), where a slight modification is required (\cite{Gaunt,BeBeJa}).

The basic rule in constructing intersections of $N>2$ fundamental objects
 is, that each pair among the $N$ objects must be one of the above pairs.
 This leads to configurations with a maximum of nine branes \cite{mult}.
 In the next section, we will discuss the role of  supersymmetry in 
 obtaining multiple intersections.

\vspace{3mm}
\noindent{\bf 3.\ Supersymmetry} 
\vspace{3mm}

The BPS $M2$-and $M5$-brane each preserves 1/2 of the $D=11$ supersymmetry.
 The supersymmetry transformation of the gravitino reads:
\be
\label{dpsi}
  \delta\psi_{\mu} = 
  \partial_\mu\epsilon - \tfrac{1}{4}\omega_\mu{}^{ab}\epsilon
   - \tfrac{i}{576}\left(\Gamma_\mu\Gamma^{abcd}-3\,\Gamma^{abcd}\Gamma_\mu
   \right)\,\epsilon\,F_{abcd}\,.
\ee
Supersymmetry is partially preserved, if the configuration is such that
 $\delta\psi_\mu$ vanishes for some $\epsilon$.
 For $M2$ and $M5$ a simple calculation leads to the following
 conditions:
\bea
  M2\ :\ \ \epsilon &=& H^{-1/6}\eta,\ \eta\ {\rm constant\ with}\ 
       P_2\eta=\eta,\ {\rm where}\ P_2 = i\Gamma^{012}\,, 
 \label{M2susy}\\
  M5\ :\ \ \epsilon &=& H^{-1/12}\eta,\ \eta\ {\rm constant\ with}\ 
       P_5\eta=\eta,\ {\rm where}\ P_5 = \Gamma^{012345}\,.
\label{M5susy}
\eea
So $\eta$ is algebraically restricted by a product of $\Gamma$-matrices
 corresponding to the worldvolume directions.
 
Given the supersymmetry preserving conditions (\ref{M2susy}, \ref{M5susy}), the
 obvious question is how to formulate the preservation of supersymmetry
 for pairs of $M$-branes. If $\eta$ must satisfy two conditions, then 
 compatibility requires that the corresponing $P_p$ must {\it commute}. 
 For
 a pair consisting of a $p_1$ and a $p_2$ brane, 
 intersecting over a common worldvolume of dimension
 $d_{12}+1$, one can derive the following rule:

\begin{itemize}
\item If both $p_1$ and $p_2$ are even, $d_{12}$ must be even,
 otherwise $d_{12}$ must be odd. 
\end{itemize}

\noindent Such a pair will preserve 1/4 of the $D=11$
 supersymmetry.
 For $M_2$ and $M_5$ this condition leads precisely to the four 
 possibilities given in (\ref{M2M2}-\ref{M5M58}).

Once intersections of three or more  fundamental branes
 have been obtained, there is a simple method to add additional
 branes which do not lead to further supersymmetry breaking.
 Consider a triple $p_1$, $p_2$ and  $p_3$ satisfying the
 above conditions, i.e., such that the $P_{p_i}$ commute.
 Then the product $P_{p_4}\equiv P_{p_1}P_{p_2}P_{p_3}$ clearly
 commutes with each $P_i$, and a brane with spatial extension
 $p_4$ can be added to the configuration. Note that this
 calculation also determines the orientation of the $p_4$-brane.

For any allowed triple of $M2$ and $M5$, one finds that $p_4$,
 calculated as above, is always one of the numbers $1,2,5,6,9$,
 i.e., $p_4$ is of the form $4k+1$ or $4k+2$. More precisely,
 we find the following: Let $p_1$, $p_2$ and $p_3$ form an intersecting
 triple with $1/8$ supersymmetry, then
\begin{itemize}
\item If either one or three $p_i$ are of the form $4k+1$, then
 so is $p_4$, otherwise $p_4$ is of the form $4k+2$.
\end{itemize}
It now becomes interesting to extend the intersecting pairs of Section 2 
 to the case of $M$-branes with spatial dimensions $1,2,5,6,9$. As we have 
 seen above, the allowed pairs are determined by supersymmetry. The result 
 is given in the Table 1. In this table we have left out intersections of the
 form $(p|p,p)$, where the two intersecting branes overlap completely.
 These are still expressed in terms of a single harmonic function
 and preserve 1/2 of supersymmetry. In the table
 the numbers $d_{12}$, $p_1$ and $p_2$ are therefore restricted by
 $d_{12} < {\rm max}(p_1,p_2)$. The fact that 
 the configuration must fit in ten spatial dimensions
 implies $p_1+p_2-d_{12}\le 10$.

\vspace{3mm}
\be
\begin{array}{|c||c|c|c|c|c|}
\hline
\mbox{$p_i$}&{1}&{2}&{5}&{6}&{9}\\
\hline
\hline
 1&   -   &   1    &   1    &    1    &    1   \\
\hline
 2&   1   &   0    &   1    &   0,2   &    1   \\
\hline 
 5&   1   &   1    &  1,3   &   1,3   &    5   \\
\hline
 6&   1   &  0,2   &  1,3   &   2,4   &    5   \\ 
\hline 
 9&   1   &   1    &   5    &    5    &    -   \\ 
\hline
\end{array}
\nonumber
\ee

\vspace{3mm}
\noindent Table 1.\ {\bf Basic pair intersections $(d_{12}|p_1,p_2)$ 
 in $D=11$.}
 The table indicates the possible values of $d_{12}$ for each pair
 $p_1$ and $p_2$. The 2- and 5-branes are discussed 
 in Section 2, the nature of
 1-, 6- and 9-branes in Section 4.
\vspace{5mm}

We have seen that supersymetry determines the pair intersections, and
 is helpful in obtaining, for a given configuration, an additional
 brane which does not lead to further supersymmetry breaking. For the
 last point we used triple configurations with 1/8 supersymmetry. 
 A further use of supersymmetry arises for the pair intersections themselves.
 Consider a pair $(d_{12}|p_1,p_2)$. By taking the product of
 $P_{p_1}$ and $P_{p_2}$ we obtain a matrix $\Gamma^{(p_1+p_2-2d_{12})}$,
 where $(p)$ stands for a set of $p$ 
 spatial indices. The indices correspond to the relative
 transverse coordinates of the pair. This matrix does not involve
 $\Gamma^0$, so the worldvolume is  spacelike and cannot be used to
 define an additional brane. But in $D=11$ the matrix $i\Gamma^{012\ldots 10}=
 1$. Therefore 
 $\Gamma^{(p_1+p_2-2d_{12})} = i\Gamma^{0(10-p_1-p_2+2d_{12})}$, which
 does define a suitable worldvolume. Note that if $p_1$ and $p_2$ 
 are both of the form $4k+1$ or $4k+2$, then so is $10-p_1-p_2+2d_{12}$.
 In this way we can obtain configurations of three branes with 1/4
 supersymmetry, which have no overall transverse directions.
 However, one has to be careful with the way the
 harmonic functions are allowed to depend on the coordinates. Following
 the rules for intersecting pairs, one finds that only in a few
 cases a nontrivial solution arises. There is only one
 example involving only $M2$ and $M5$.
 This arises from the pair $(1|5,5)$ (see (\ref{M5M58})), to which we
 can add an $M2$, such that the triplet has a common string direction
 (see also \cite{Tseyt3,Gaunt2}).

\vspace{3mm} 
\noindent{\bf 4.\ The 1-, 6- and 9-brane}
\vspace{3mm} 

In Table 1 we find the pairs (\ref{M2M2}-\ref{M5M58}) as a subset.
 Now we must discuss the nature of the branes of extension $1,6$ and $9$.
 For the first two cases we have obvious candidates. The $M1$-brane
 can be interpreted as the Brinkmann wave in $D=11$:
\be
  \label{wave}
ds^2= (2-H)dt^2 -Hdz^2 + 2(1-H)\,dtdz - (dx_2^2 + ... +dx_{10}^2), 
\ee
where $H$ is a harmonic function in the variables $t+z, 
 x_2,\ldots, x_{10}$. Its interpretation as an $M1$-brane makes sense, 
 since it indeed preserves 1/2 supersymmetry, and its direct
 dimensional reduction to $D=10$ gives the fundamental string solution.
 The double dimensional reduction gives the $D0$-brane in $D=10$.
 
Also the $M6$-brane allows a natural interpretation. It must be the
 Kaluza-Klein monopole \cite{Sorkin}, with metric ($i=1,2,3$)
\be
\label{monopole}
ds^2 = dt^2 - dx_1^2 - ... -dx_{6}^2 -H^{-1}(dz + A_i dy_i)^2 
      - H dy_i^2\,,  
\ee
where $H$ and $A_i$ depend on $y_i$, and 
the relation between $H$ and $A_i$ is
\be
F_{ij} \equiv \partial_i A_j - \partial_j A_i =  \epsilon_{ijk}\partial_k H\,.
\ee
Direct dimensional reduction to $D=10$ gives a $D6$-brane, double
 dimensional reduction the solitonic fivebrane in $D=10$.
 Recently we have extended our results on $M2$- and $M5$-branes \cite{mult}
 to include also the wave (\ref{wave}) and the monopole (\ref{monopole})
 \cite{usnew}.
 Interestingly, the intersections of pairs of waves and monopoles 
 with $M2$ and $M5$, and with themselves, are precisely as given in
 Table 1. This, and the results on multiple
 intersections \cite{mult}, 
 gives us some confidence that supersymmetry may indeed be 
 used to predict the allowed configurations of intersecting branes. 
 According to this point of view,
 the construction of a multiple intersections involving $N$ 
 basic objects is the same  as the construction of $N$ commuting
 matrices $\Gamma^{0(p_i)}$, $i=1,\ldots N$, where $(p_i)$ denotes
 the spatial orientation of the worldvolume of the $p_i$-brane.

There is no known 9-brane solution of $D=11$ supergravity. Nevertheless,
 the above results indicate that we should seriously consider the
 existence of such an object\footnote{The $D=11$ 9-brane has
 been discussed before. See remarks in \cite{Be2,Howe,Papa2,Pol1,Duff}.}.
 There are also other indications
 that a 9-brane should exist. In $D=10$ there is an $D8$-brane
 solution \cite{Pol2,Be2}, and, according to the $M$-theory interpretation
 of string theories, it should have
 an eleven-dimensional counterpart. However, the $D8$-brane requires
 the  massive extension of $D=10$ IIA supergravity \cite{Romans}, which
 we do not know how to lift to $D=11$. 

Our analysis does not tell us what the conjectured 9-brane solution is.
 But, assuming that it preserves 1/2 supersymmetry, and that
 the condition of preservation of supersymmetry is of the
 standard form, its pair intersections with the known solutions of
 $D=11$ supergravity are determined (see Table 1). 
 For instance, this analysis tells
 us that the 9-brane can occur in configurations of $n$ $M5$-branes 
 for $n\le 7$. Such configurations would reduce in $D=10$ to
 an intersection of $n$ $D4$-branes with the $D8$-brane, which is known to
 be a solution of massive $D=10$ IIA supergravity.

\vspace{3mm}
\noindent{\bf Acknowledgements} 
\vspace{3mm}

It is a pleasure to thank the Organising Comittee of this meeting for
 their invitation, and the participants from the Ukraine and abroad
 for the pleasant atmosphere during this meeting. The work 
 described here was done in collaboration with
 Eric Bergshoeff, Eduardo Eyras, Bert Janssen and
 Jan Pieter van der Schaar.
 This work is also supported  by the European Commission TMR programme 
 ERBFMRX-CT96-0045,
 in which I am associated to the University of Utrecht.

\end{document}